\documentclass[10pt,conference]{IEEEtran}
\IEEEoverridecommandlockouts
\usepackage{cite}
\usepackage{textcomp}
\usepackage{graphicx}
\usepackage[shortlabels]{enumitem}
\usepackage{balance}
\usepackage{comment}

\usepackage{url}
\usepackage{color}

\usepackage{amsmath,amsfonts,amssymb}
\usepackage{bm}
\usepackage{algorithm}
\usepackage{algpseudocode}
\usepackage{listings}
\usepackage{multirow}
\usepackage[numbers]{natbib}
\usepackage{blindtext}
\usepackage{wrapfig}
\usepackage{tcolorbox}

\def\BibTeX{{\rm B\kern-.05em{\sc i\kern-.025em b}\kern-.08em
		T\kern-.1667em\lower.7ex\hbox{E}\kern-.125emX}}
	
\newcommand{\inlinedComment}[2]
{\textcolor{#1}{\small\textbf{#2}}}
\newcommand{\n}[1]{}
\newcommand{\lx}[1]{}
\newcommand{\yu}[1]{}

\begin{document}
	
	
	\title{\huge InferCode: Self-Supervised Learning of Code Representations by Predicting Subtrees}


	\author{\IEEEauthorblockN{Nghi D. Q. Bui}
		\IEEEauthorblockA{\textit{School of Information Systems} \\
			\textit{Singapore Management University}\\
			dqnbui.2016@phdcs.smu.edu.sg}
		\and
		\IEEEauthorblockN{Yiyun Yu}
		\IEEEauthorblockA{\textit{School of Computing \& Communications} \\
			\textit{The Open University, UK}\\
			y.yu@open.ac.uk}
		\and
		\IEEEauthorblockN{Lingxiao Jiang}
		\IEEEauthorblockA{\textit{School of Information Systems} \\
			\textit{Singapore Management University}\\
			lxjiang@smu.edu.sg}
	}
	
	\maketitle
	
	\begin{abstract}
Learning code representations has found many uses in software engineering, such as code classification, code search, code comment generation, and bug prediction.
Although representations of code in tokens, syntax trees, dependency graphs, paths in trees, or the combinations of their variants have been proposed, existing learning techniques have a major limitation that these models are often trained on datasets labeled for specific downstream tasks, and the code representations may not be suitable for other tasks.
Even though some techniques generate representations from unlabeled code, they are far from satisfactory when applied to downstream tasks.
To overcome the limitation, this paper proposes {\em InferCode}, which adapts the self-supervised learning idea from natural language processing to the abstract syntax trees (ASTs) of code.
The key novelty lies in the training of code representations by predicting {\em subtrees} automatically identified from the context of ASTs.
With InferCode, subtrees in ASTs are treated as the labels for training the code representations without any human labeling effort or the overhead of expensive graph construction, and the trained representations are no longer tied to any specific downstream tasks or code units.

We have trained an instance of InferCode model using Tree-Based Convolutional Neural Network (TBCNN) as the encoder of a large set of Java code.
This pre-trained model can then be applied to downstream unsupervised tasks such as code clustering, code clone detection, cross-language code search, or be reused under a transfer learning scheme to continue training the model weights for supervised tasks such as code classification and method name prediction.
Comparing to prior techniques applied to the same downstream tasks, such as code2vec, code2seq, ASTNN, using our pre-trained InferCode model higher performance results are achieved with a significant margin for most of the tasks, including those involving different programming languages.
The implementation of InferCode and the trained embeddings are made available at the anonymous link: \url{https://github.com/ICSE21/infercode}.

	\end{abstract}
	
\section{Introduction}
\label{sec:1}

Learning code representations (a.k.a.~embeddings) and building a prediction model for programs have been found useful in many software engineering tasks, such as classifying program functionality \cite{Nix2017,dahl2013large}, code search \cite{gu2018deep,kim2018facoy,Sachdev2018}, code comment generation \cite{hu2018deep,Wan2018,Alon2019}, predicting bugs \cite{Yang2015,li2017software,zhou2019devign}, translating programs \cite{chen2018tree, Gu2017}, etc.
While offering promising performance for the tasks, the prior learning techniques have major limitations that hinder their performance and generalizability.
\begin{itemize}[nosep,leftmargin=1em]
	\item Most of the code representations and program models are trained in a (semi-)supervised learning paradigm. Human needs to manually label the data for a specific downstream task, or engineer some special intermediate representations and corresponding training techniques for the task, and the code representations are trained with respect to the specific task. Not to mention the efforts needed to provide many labels and specially engineered features, such trained code representations are specific to one particular task and may not be easily transferred to other tasks. 
	\item Even though there are techniques \cite{alon2018code2seq, Alon2019} aiming to produce code representations that are transferable to different tasks, their trained code representations are only for some fixed units of code, such as tokens, statements, and functions, and are not flexible to produce embeddings for different code units. Such techniques may miss useful information across different kinds of code units, and the trained representations may not perform well for various downstream tasks either.
	Some other techniques based on graph embeddings \cite{Narayanan2017,fang2020functional,wang2020detecting} share a similar drawback and in addition need the overheads of graph construction which may introduce inaccurate information in the graphs.
	
\end{itemize}

Such limitations have been illustrated in a recent study: 
\citet{kang2019assessing}
show that the pre-trained code2vec~\cite{Alon2019} representation does not perform well for other tasks when it was trained specifically for the method-name prediction task.

Towards addressing the limitations, {\bf the aim} of this paper is to develop a new technique for learning code representations, and it should be: (1) trainable without any manual human labeling, (2) flexible in producing embeddings for any code unit that can be parsed into syntax trees, and (3) general enough so that its trained representations for code can perform well for various downstream tasks.

We have two pillars that support the realization of our aim. One is the large amount of source code available on public code hosting platforms, such as Github, Bitbucket, Gitlab. Although the code often lacks accurate labels for downstream tasks, the syntax of the code itself can be relatively easily checked by parsers. It is desirable to leverage such unlabeled data to pre-train code representations reusable for building various program prediction models for downstream tasks. 

The second supporting pillar is the advances of self-supervised learning in the machine learning community \cite{hinton2006fast, chen2020simple, yasunaga2020graph, doersch2017multi, kolesnikov2019revisiting}.
Such techniques enable the training of neural networks without the need for human labels. Usually, a self-supervised learning technique reformulates an unsupervised learning problem as a supervised one by {\em automatically generating virtual labels from existing (unlabeled) data}.
The self-supervised task, also known as a {\em  pretext task}, guides us to a supervised loss function.
While minimizing the loss function for the pretext task, the technique can also produce intermediate representations for the data corresponding to the virtual label. Because the pretext task can be trained using any data, it is expected that such representations can carry good information of diverse data and be beneficial to a variety of downstream tasks. 
This notion of self-supervised learning is very suitable for our aim.
Little effort has been invested in the literature to exploit the uses of self-supervised learning for code representation learning. Although some recent work, such as \cite{yasunaga2020graph}, presents a self-supervised learning paradigm for program repair, it is designed specifically for the specific task.

\begin{figure}[!t]
	\centering
	\includegraphics[scale=0.18]{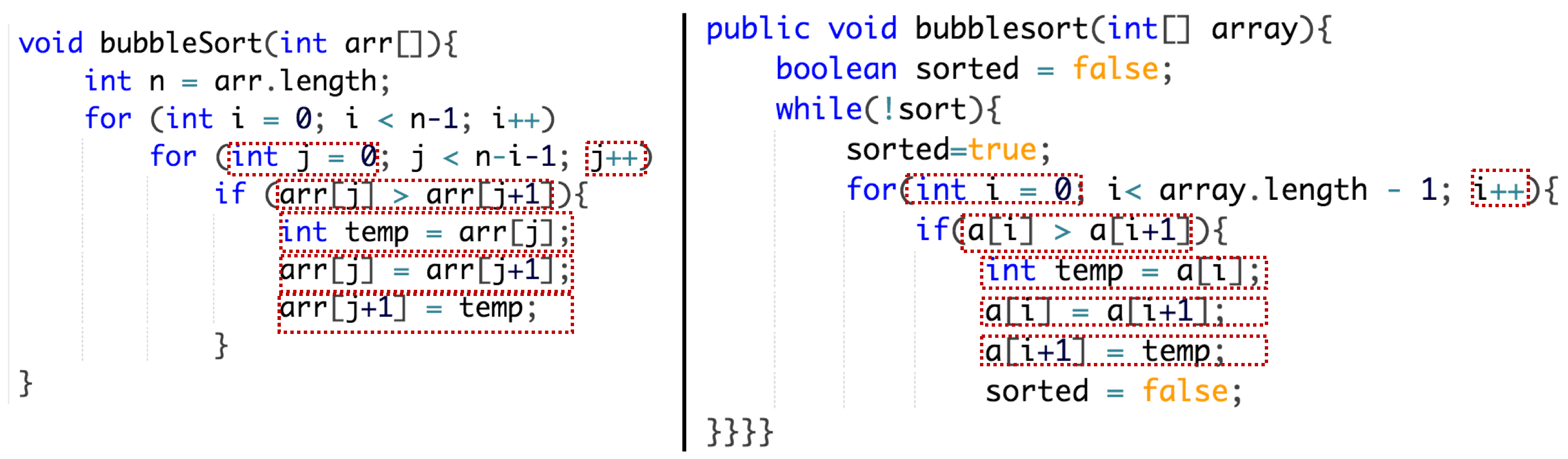}
	\vspace*{-20pt}
	\caption{Example of two code snippets that implement bubble sort in Java that share similar fine-grained code elements.}
	\label{fig:intuition}
\end{figure}

Our key idea is thus to train a pretext task suitable for any source code. Different from self-supervised learning in natural language processing and visual learning areas that use words or object regions as labels, we utilize the fact that it is relatively easy to obtain the abstract syntax tree (AST) of any syntactically valid code snippet via parsers and it is also easy to identify all the subtrees in ASTs, and automatically use each subtree as the label for the pretext task to predict the probability of the subtree appearing in a particular AST.\footnote{An underlying assumption, for such trained representations to capture code meanings, is that code snippets with the same semantics should involve some syntactically similar code elements. Even though two pieces of code implementing the same functionality can be syntactically very different, there could still be some fine-grained elements in the code or other pieces of code that use these two that are syntactically similar, when the code base is large.}
Fig.~\ref{fig:intuition} shows an example of this intuition. The two code snippets implement the same functionality, bubble sort. If we view these two code snippets as two ASTs, there are many similar subtrees between these two AST. For example, the subtree that represents the conditional expression \texttt{arr[j] > arr[j+1]} of the left snippets is similar to \texttt{arr[i] > arr[i+1]} although the textual information is quite different. This means that if we can exploit such information, we do not need any label to build a representation learning model for source code.
Also different from recent uses of neural document embedding models (e.g., doc2vec \cite{mikolov2013distributed,le2014distributed}) for source code (e.g., \cite{wei2017supervised,Ingram2018,Broggi2018,Aman2018,chen2019literature,akbar2019scor}),
our technique learns subtrees in ASTs without the overheads and accuracy losses of constructing customized graphs,
while they learn mostly code tokens and node types, although we are all inspired by the same idea of doc2vec.
We also provide an alternative to graph-based \cite{Narayanan2017,tufano2018deep} or execution traces-based \cite{wang2019learning} embedding techniques as we believe ASTs are more readily available for all kinds of programming languages and may have contained all the code information (although some are hidden).


Based on the key idea, we propose \textbf{InferCode}, a self-supervised learning technique for source code by predicting syntax subtrees.
As far as we know, we are the first to apply the notation of self-supervised learning to syntax subtrees and can produce code representations for any syntactically valid code snippet without the need of human labeling.
\begin{itemize}[leftmargin=*]
	\item InferCode can serve as an \textit{encoder} that maps any parsable code snippet
	into a vector representation (embedding), and this vector can be used for various downstream tasks, such as code clustering, clone detection, and code search.
	\item InferCode can serve as a pre-trained model and its weights can be reused in downstream training of the models for supervised learning tasks, which can speed up the training and alleviate the issue of lacking data for a particular task.
	\item We implement InferCode on top of the ASTs produced by SrcML~\cite{collard2013srcml}. It provides a combined vocabulary of AST node types for multiple languages (e.g., Java, C, C++, C\#), which implies that our InferCode can be polyglot, producing code representations suitable for tasks involving different languages, such as cross-language code search, as long as the ASTs for a code snippet can be recognized by SrcML.
\end{itemize}

We have trained an instance of InferCode based on a large set of Java code and evaluated the usefulness of the pretrained code representations in five downstream tasks, three of which are unsupervised (code clustering, code clone detection via similarity measurement, cross-language code search, two are supervised (code classification and method name prediction).
For the three unsupervised tasks, we utilize the vectors produce by InferCode and different vector similarity metrics to achieve the goal of each task: For \textit{code clustering}, our results using InferCode outperform the best baseline (Code2vec) by 12\% in term of Adjusted Rand Index; For \textit{code clone detection}, our results outperform the best baseline (Code2vec) by 15\% in term of F1 score; For \textit{cross-language code search}, our results outperform the best baseline (CLIR) on 13\% (on average for multiple languages setting) in term of Mean Reciprocal Rank.
For the two supervised tasks, we utilize the weights of the pre-trained model from InferCode to fine-tune the specific prediction model for each task: 
our results using the fine-tuning process increases the performance of TBCNN for \textit{code classification} by 4\% in term of accuracy, which is comparable to ASTNN, the state-of-the-art model for code classification, , and increase the performance TBCNN for ~\textit{method name prediction} by 8\%, which is comparable to code2seq, a state-of-the-art model for method name prediction.

\section{Related Work}\label{sec:related}
\textbf{Self-Supervised Learning} has made great progress recently for visual data~\cite{mahendran2018cross, gidaris2018unsupervised, zhang2016colorful,korbar2018cooperative,kim2019self,fernando2017self}: \citet{gidaris2018unsupervised} proposed a method to generate different viewpoints of an image by a number of rotations on certain degrees at random and formulate the learning part as a multi-class classification problem over the rotations. This pretext task drives the model to learn semantic concepts of objects as the parameters of the CNN image encoder; \citet{zhang2016colorful} proposed to use colorization as the pretext task by giving colours to a grayscale input image to map this image to a distribution over quantized color value outputs.

There has been tremendous effort to explore self-supervised learning in Natural Language Processing research for quite a while~\cite{mikolov2013distributed, le2014distributed, kiros2015skip, devlin2018bert}. Word2vec~\cite{mikolov2013distributed} is a form of self-supervised learning, which aims to learn good representation for words by taking a small chunk of the text of certain window size. Doc2vec~\cite{le2014distributed} shares the same principle with word2vec which aims to use a document to predict the words inside it so that similar documents will have similar embeddings; Skip-thought vectors~\cite{kiros2015skip} builds a language model by predicting the neighbouring sentences of a center sentence; BERT~\cite{devlin2018bert} advances language models by masking the words in a text randomly in order to predict them. 

\textbf{Deep Learning Models of Code}: There has been a huge interest in applying deep learning techniques for software engineering tasks such as program functionality classification~\cite{mou2016convolutional,zhang2019novel}, bug localization~\cite{pradel2018deepbugs, gupta2019neural}, function name prediction~\cite{fernandes2018structured}, code clone detection~\cite{zhang2019novel}, program refactoring~\cite{hu2018deep}, program translation~\cite{chen2018tree}, and code synthesis~\cite{brockschmidt2018generative}.
\citet{Allamanis2018} extend ASTs to graphs by adding a variety of code dependencies as edges among tree nodes, intended to represent code semantics, and apply Gated Graph Neural Networks (GGNN)~\cite{Li2016} to learn the graphs;
Code2vec~\cite{Alon2019}, Code2seq~\cite{alon2018code2seq}, and ASTNN~\cite{zhang2019novel} are designed based on splitting ASTs into smaller ones, either as a bag of path-contexts or as flattened subtrees representing individual statements. They use various kinds of Recurrent Neural Network (RNN) to learn such code representations. Unfortunately, there is little effort that invests to design the source code model with unlabeled data: \citet{yasunaga2020graph} presents a self-supervised learning paradigm for program repair; Survey on code embeddings~\cite{Ingram2018,chen2019literature} presents evidence to show that there is a strong need to alleviate the requirement of labeled data for code modeling and encourage the community to invest more effort in the methods on learning source code with unlabeled data. 

Our approach differs from existing ways to reuse the pre-trained code learning model: \citet{kang2019assessing} reuse the token embeddings from Code2vec for downstream tasks only to find that lower performance than simpler word embedding methods like Word2vec. In contrast, we use the weights of the pretrained model and the code vector $\vec{v}$ produced by the encoder instead of the token embeddings. 
\n{more things about this}
\lx{add a few sentences about graph2vec etc. too}

\section{Preliminaries}\label{sec:background}
\begin{figure*}[t]
	\centering
	\includegraphics[scale=0.36]{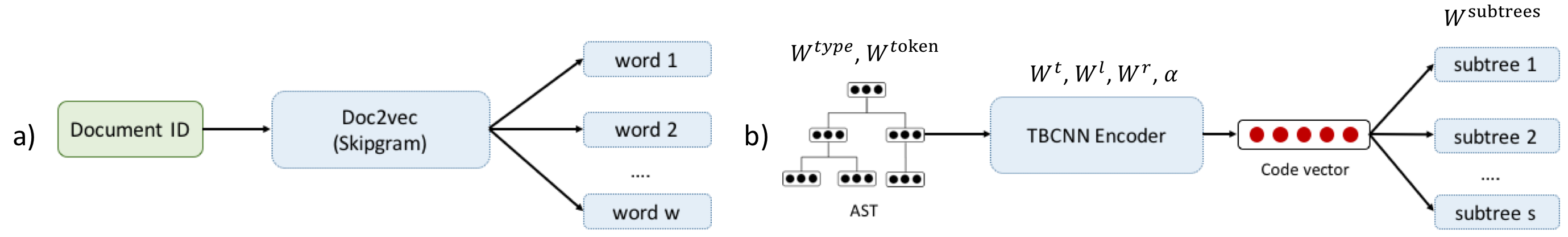}
	\caption{a) Doc2vec's skipgram model - Given a document \textit{d}, it samples \textit{c} words and considers them as co-occurring in the same context of \textit{d} to learn \textit{d}'s representation; (b) InferCode - Given an AST $T$, it samples \textit{s} subtrees from \textit{T} and uses them as the context to learn $T$'s representation.}
	\label{fig:overview}
\end{figure*}

\subsection{Source Code Representation Learning}
\label{sec:prem:codelearning}
Source code representation learning usually contains the following two phases: (1) represent a code snippet into an intermediate representation (IR), such as token streams, ASTs, AST paths or graphs; and (2) design a neural network suitable to process such intermediate representations.
The neural network can also be called as an {\em encoder}. The encoder receives the code IR and maps it into a code vector embedding $\vec{v}$ (usually a combination of various kinds of code elements), then $\vec{v}$ can be fed into the next layer(s) of a learning system and trained for an objective function of the specific task of the learning system.
For example, in Code2vec \cite{Alon2019}, $\vec{v}$ is a combination of different AST paths. In GGNN~\cite{Allamanis2018} or TBCNN~\cite{mou2016convolutional}, $\vec{v}$ is a combination of AST nodes.
A trained model, either on supervised learning or self-supervised learning task can produce the $\vec{v}$.
In our work, we will evaluate how a $\vec{v}$ trained on a self-supervised learning objective over a large set of unlabeled data can be useful for different tasks. 


\subsection{Neural Document Embedding Models}
Doc2vec \cite{le2014distributed} is an extension to word2vec \cite{mikolov2013distributed}.
Doc2vec uses an instance of the skip-gram model called paragraph vector-distributed bag of words (interchangeably referred as doc2vec skip-gram) that is capable of learning representations of word sequences of arbitrary lengths, such as sentences, paragraphs and even whole large documents.
More specifically, given a set of documents $\{d_1,d_2,...d_n\}$ and a sequence of words $\{ ..., w_{ij}, ... \}$ sampled from the document $d_i$, skip-gram learns a $D$-dimensional embeddings of the document $d_i$ and each word $w_{ij}$ sampled, i.e., $\vec {v}_i, \vec{v}_{ij} \in \mathbb{R}^D$, respectively.
The model works by considering a word $w_{ij}$ to be occurring in the context of document $d_i$ and tries to maximize the following log likelihood: 
$
\label{eq:doc2vec}
\sum_{j} log \ Pr (w_{ij}  | d_i)
$, 
where the probability  $ Pr (w_{ij}  | d_i)$ is defined as
$
\label{eq:doc2vec_softmax}
\frac {exp(\vec{v}_i \cdot \vec{v}_{ij})} {\sum_{w \in \mathcal{V}} exp(\vec{v}_i \cdot \vec{w})}
$,
where $\mathcal{V}$ is the vocabulary of all the words across all documents.  

In this paper, we consider ASTs analogous to documents and subtrees in ASTs analogous to words in documents, and adapt the idea of document embedding to learn embeddings of ASTs of any size by using an encoder that can encode ASTs of any parsable code snippet.


\subsection{Self-supervised Learning Formulation}

The goal of self-supervised learning is to train an encoder $E$ such that $E$ can map an object into a vector representation (embedding). In our case, the embedding $\vec{v}$ is for the AST representation $T$ of a code snippet $C$. 
Training the encoder $E$ is to learn its parameters (or weights) so that $E$ is able to produce the embeddings for the code snippets such that the vectors for the snippets having similar syntactical and semantic information will be close in the vector space. In visual learning, Convolutional Neural Networks are usually chosen as the encoder for images. In NLP, Recurrent Neural Networks, or recently, BERT, is used as the encoder for text sequences.
In our case, we choose Tree-based CNN as the source code encoder as it has been successfully used before \cite{mou2015discriminative,mou2016convolutional,bui2017cross,yu2019neural} and justified further in Section~\ref{sec:discussion}.

Given a dataset $X$, for each data $X_{i}$ in X, there is a corresponding pseudo label $P_{i}$ automatically generated for a predefined pretext task without involving any human annotation. 
Given a set of $n$ training data $D = \{P_{i}\}_{i=1}^n$, the aim is to minimize the loss function:
$
loss(D) = \frac{1}{n}\sum_{i=1}^{n} loss({X_{i}},P_{i})
$.
We can easily identify subtrees in ASTs as the pseudo labels $P$ automatically without human annotations so that our learning technique can be self-supervised.

%

\section{Approach Details}\label{sec:approach}
\subsection{Overview}

Figure \ref{fig:overview} presents a high-level view of our InferCode approach as an analogy to Doc2vec by treating an entire AST as a document and treating the subtrees as words.  
Given a set of ASTs $\{T_1,T_2,...T_n\},$ and a set of all subtrees $\{ ..., T_{ij}, ... \}$ of $T_i$, we represent $T_i, T_{ij}$ by $D$-dimensional embedding vectors $\vec{v}_i, \vec{v}_{ij} \in \mathbb{R}^{D}$, respectively.
By considering a subtree $T_{ij} \in T_i$ to be occurring in the context of the AST $T_i$, we aim to maximize the following logarithmic likelihood: 
$
\sum_{j} log \ Pr (T_{ij}  | T_{i})
$.

Different from doc2vec, InferCode does not query the embedding vectors directly from an embedding matrix for the whole documents;
instead, we first encode the entire AST to obtain the  $\vec{v}_{i}$, then use it to predict the subtrees.
The steps of our technique are as follows:
\begin{itemize}[nosep,leftmargin=*]
	\item For each AST in our dataset, we identify a set of subtrees, and all of the subtrees are accumulated into a vocabulary of subtrees (Section~\ref{sec:generate_subtrees});
	\item We feed an AST into a Tree-Based CNN (TBCNN) encoder to produce a code vector $\vec{v}_{i}$. Then $\vec{v}_{i}$ is used to predict the subtrees identified in the previous step;
	\item After the encoder has been trained, we can then use it as the pretrained model for downstream tasks.
\end{itemize}

\subsection{Process to Identify Subtrees}
\label{sec:generate_subtrees}
\begin{figure}[!h]
	\centering
	\includegraphics[scale=0.17]{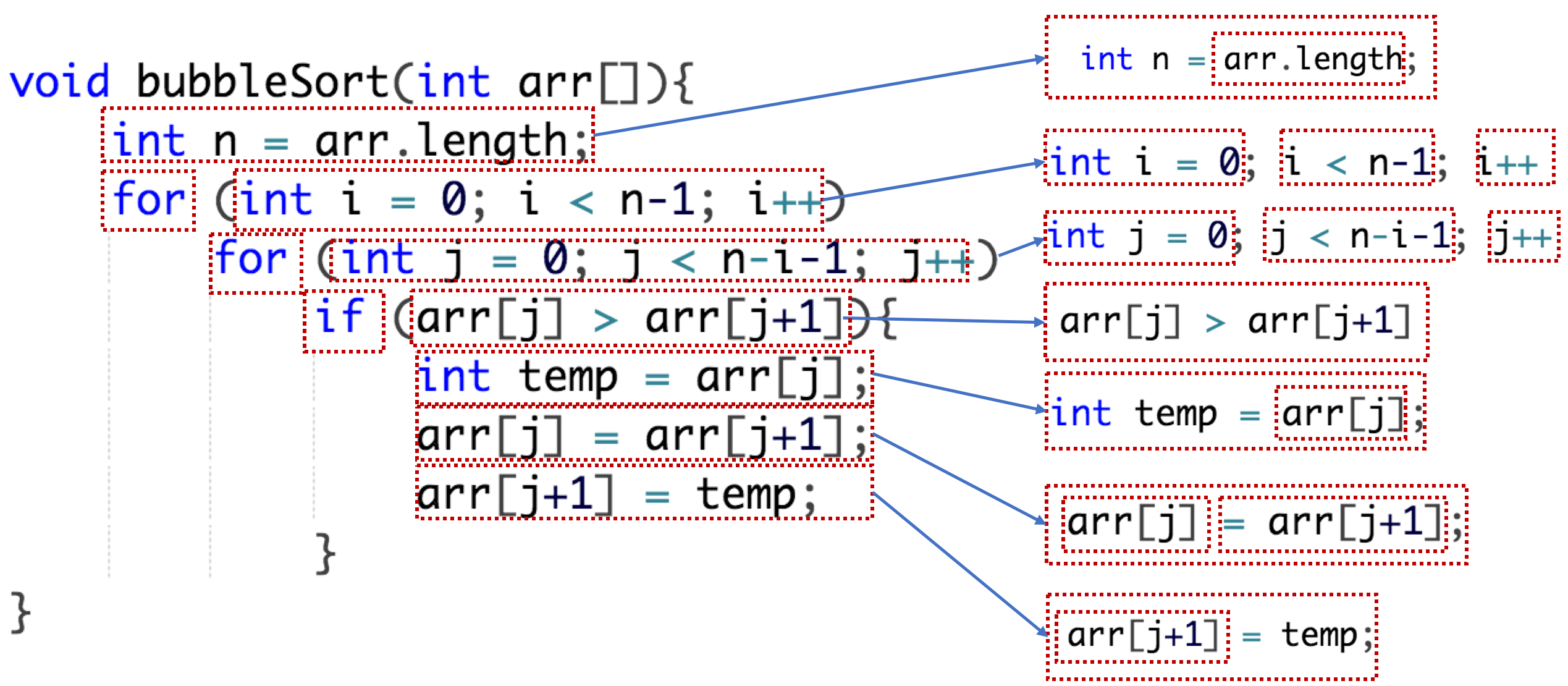}
	\vspace*{-5pt}
	\caption{Example to generate subtrees from a code snippet}
	\label{fig:generate_subtrees}
\end{figure}

By traversing an AST, every visited node satisfying a certain condition, e.g., of the type \texttt{expr}, leads to a subtree rooted at the visited node. In our experiments, we chose to select the subtrees whose root nodes are of the types \{\texttt{expr\_stmt}, \texttt{decl\_stmt}, \texttt{expr}, \texttt{condition}\},
We consider these relatively fine-grained code elements because they are usually meaningful but yet still small enough to be considered as the frequent ``words'' in the vocabulary of subtrees from a large code base. Such small code elements often have similar meaning when their syntactical structure is similar even though their textual appearance may be different (due to different identifier names, such as \texttt{int n = arr.length} versus \texttt{int m = x.length}). In addition, we also consider nodes that represent for a single keyword, such as \texttt{if, for, while}. Noted that these nodes can be seen as the sutrees with size = 1.

We do not consider too coarse-grained subtrees, such as the whole \texttt{if, while, for} statements, as those subtrees are often big so that (1) each of them, as an individual vocabulary word, may appear too infrequent in the code base for the encoder to learn a meaningful representation for it directly; (2) syntactical differences among the big subtrees do not necessarily mean the corresponding code has different meanings, while the encoder may have harder time to recognize the semantic similarity among them. 

Figure~\ref{fig:generate_subtrees} shows a sample bubble sort code snippet written in Java
and the identified subtrees on the right side.
This snippet is parsed into an AST, and certain subtrees are identified automatically.
For example, the statement \texttt{int n = arr.length} contains an expression \texttt{arr.length}.
Both 
\texttt{int n = arr.length} and \texttt{arr.length} are identified.

\subsection{Learning Source Code Representation}
Once we have the subtrees, we can use it to learn the source code encoder under a self-supervision mechanism.  Here we choose TBCNN~\cite{mou2016convolutional} as the source code encoder. There are two major differences between our TBCNN and the original design of \cite{mou2016convolutional}: we include the textual information into the node initialization embedding instead of using only the type information, and we replace the dynamic max pooling with an attention mechanism to combine node embeddings. Figure~\ref{fig:tbcnn} shows an overview of the workflow of the TBCNN with the modifications that we made.
There are 3 steps to learn the weights of the encoder, which can be described as:

\begin{figure}[!t]
	\centering
	\includegraphics[scale=0.25]{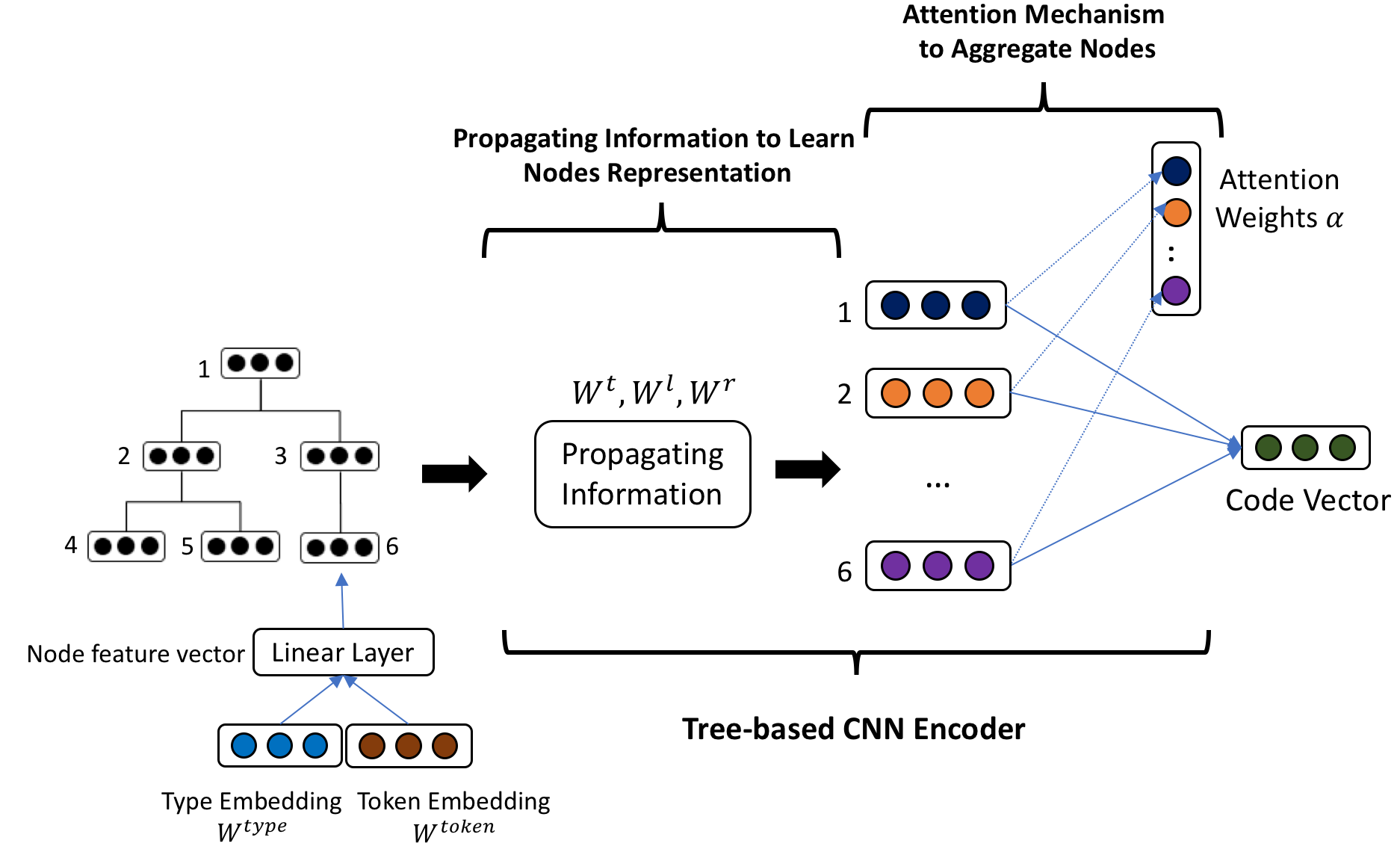}
	
	\caption{Workflow of Tree-based Convolutional Neural Network~\cite{mou2016convolutional} with 2 modifications: (1)including the token information to initialize the node vector; and (2) use the attention mechanism to aggregate nodes information}
	\label{fig:tbcnn}
\end{figure}
\begin{itemize}[leftmargin=*]
	\item \textbf{Learning Nodes Representation}: This step is to learn the representation of the node of the input AST $T$. The information of the tree will propagate from bottom to top, i.e., a parent node will accumulate the information of its descendant in the AST. After the accumulation step, each node will contain the information of its descendants.
	\item \textbf{Aggregating Nodes Information}: Since we want to represent the AST representation of the code snippet into a fixed dimension vector $\vec{v}$, we need to combine all of the node embeddings into one fixed single embedding. We use the attention layer for this purpose.
	\item \textbf{Predicting Subtrees}: Once we have the $v_{C}$, we use it to predict the subtrees extracted from $T$. Intuitively, this process is similar to Eq. \ref{eq:doc2vec_softmax}, where the task is to predict the probability of a subtree given the embedding $v_{C}$. 
\end{itemize}

\subsubsection{Learning Nodes Representation with TBCNN}
We briefly introduce the Tree-based Convolutional Neural Networks (TBCNN, \cite{mou2016convolutional}) for processing AST inputs.

A tree $T = (V, E, X)$ consists of a set of nodes $V$, a set of node features $X$, and a set of edges $E$. An edge in a tree connects a node and its children. 
Each node in an AST also contains its corresponding texts (or tokens) and its type (e.g., operator types, statement types, function types, etc.) from the underlying code.
Initially, we annotate each node $v \in V$ with a $D$-dimensional real-valued vector $\vec{x_{v}} \in \mathbb{R}^D$ representing the features of the node. We associate every node $v$ with a hidden state vector $\vec{h_{v}}$, initialized from the feature embedding $\vec{x_{v}}$. In \cite{mou2016convolutional}, the node is initialized only with the type embedding. In our case, we initialize the node with a fusion of the embeddings of its texts and through a linear layer.
The embedding matrices for the texts and types are learn-able in the whole model training pipeline, formally defined as $\mathbf{W}^{type}$ and $\mathbf{W}^{token}$, respectively.

In TBCNN, a convolution window over an AST is emulated via a binary tree, where the weight matrix for each node is a weighted sum of three fixed matrices $\mathbf{W}^t$, $\mathbf{W}^l$, $\mathbf{W}^r \in \mathbb{R}^{D \times D}$
(each of which is the weight for the ``top'', ``left'', and ``right'' node respectively)
and a bias term $\mathbf{b} \in \mathbb{R}^{D}$
Hence, for a convolutional window of depth $d$ in the original AST with $K = 2^d -1$ nodes (including the parent nodes) belong to that window with vectors $[\mathbf{x}_1, ... , \mathbf{x}_{K}]$, where $\mathbf{x_{i}}\in \mathbb{R}^{D}$,
the convolutional output $\mathbf{y}$ of that window can be defined as 
$
\mathbf{y} = tanh(\sum_{i=1}^{K}[\eta_{i}^{t}\mathbf{W}^{t}+\eta_{i}^{l}\mathbf{W}^{l}+\eta^{r}_{i}\mathbf{W}^{r}]\mathbf{x}_{i}+\mathbf{b})
$, 
where $\eta_{i}^{t},\eta_{i}^{l},\eta_{i}^{r}$ are weights calculated corresponding to the depth and the position of the nodes.

\subsubsection{Attention Mechanism to Aggregate Nodes}
\label{sec:attention}
After the nodes representation has been learned, we need an aggregation method to combine all the nodes in to one fixed embedding that represent for the code snippet. \citet{mou2016convolutional} use max pooling to combine the nodes. However, max pooling may discard a lot of important information, so we replace it with the attention mechanism to aggregate nodes.
Formally, an attention vector $\vec{a} \in \mathbb{R}^{D}$ is initialised randomly and learned simultaneously with updates of the networks. Given $n$ node state vectors: $\{\vec{h_{1}}, ..., \vec{h_{n}}\}$, the attention weight $\alpha_{i}$ of each $\vec{h_{i}}$ is computed as the normalised inner product between the node state vector and the global attention vector:
$
\alpha_i = \frac{\exp(\vec{{h_{i}}}^{T}\cdot \vec{a})}{\sum_{j=1}^{n}\exp(\vec{h_{j}}^{T}\cdot \vec{a})}
$.
The exponents in this equation are used to make the attention weights positive, and they are divided by their sum to have a max value of $1$, as done by a standard softmax function.

The aggregated code vector $\vec{v}\in \mathbb{R}^{D}$ represents the whole code snippet. It is a linear combination of the node state vectors $\{\vec{h_{1}}, ..., \vec{h_{n}}\}$ weighted by their attention scores:
\begin{equation}
\vec{v} = \sum_{i=1}^{n}\alpha_i \cdot \vec{h_i}
\label{codevector}
\end{equation}

\subsubsection{Predicting Subtrees}

From the process to extract the subtrees, we have a vocabulary of all subtrees from our training dataset. The embeddings of subtrees are learn-able parameters, formally defined as $\mathbf{W}^{subtrees} \in \mathbb{R}^{\left|L\right|\times D}$, where $L$ is the set of subtrees extracted from the training corpus. The embedding of $subtrees_i$ is row $i$ of $\mathbf{W}^{subtrees}$. The predicted distribution of the model $q\left (l\right)$ is computed as the (softmax-normalized) dot product between the code vector $\vec{v}$ and each of the subtree embeddings:
$
\small
for\, l_{i}\in L:\, q\left(l_i\right)=\frac{\exp(\vec{v}^T\cdot \mathbf{W}^{subtrees}_{i})}{\sum_{l_{j}\in L}\exp(\vec{v}^T\cdot \mathbf{W}^{subtrees}_{i})}
$
where $q\left(l_i\right)$ is the normalized dot product between the vector of $l_i$ and the code vector $\vec{v}$, i.e., the probability that a subtrees $l_i$ appears given code snippet $\boldsymbol{C}$. This is aligned with Eq. \ref{eq:doc2vec_softmax} in Doc2vec to predict the likelihood of a word given a document. 

Totally, we need to learn these parameters of InferCode: $\mathbf{W}^{type}$, $\mathbf{W}^{token}$, $\mathbf{W}^t$, $\mathbf{W}^l$, $\mathbf{W}^r \in \mathbb{R}^{D \times D}, a \in \mathbb{R}^{D},\mathbf{W}^{subtrees} \in \mathbb{R}^{\left|L\right|\times D}$. 

\subsection{Usage of the Model after Training}
\label{sec:modelusage}
We have presented the pipeline to train InferCode by predicting subtrees as the labels.
Note that in self-supervised learning, one does not usually care about the performance of the pretext task.
Instead, we care about the weights that have been learned and the ability of the model to generate the embeddings.
The trained TBCNN encoder of InferCode can be used to produce an embedding vector $\vec{v}$ for any parsable code snippet by (1) parsing the code into an AST and (2) feeding the AST through the encoding step presented in Figure~\ref{fig:tbcnn} to get the vector.
The weights in the trained model can also be used for the prediction models in downstream supervised learning tasks to save training costs and potentially improve their prediction accuracies.
We illustrate the usages in next sections.

\section{Use Cases}\label{sec:use_case}
 In this section, we briefly describe how InferCode can be adapted into 5 different downstream tasks.

\subsection{Code Embedding Vectors for Unsupervised Tasks} 
\subsubsection{Code Clustering}
Code clustering task is to put similar code snippets automatically into the same cluster without any supervision.
Given the code vectors $\vec{v}$ produced by the pre-trained InferCode for any code snippets, we can realize the task by defining a similarity metric based on Euclidean distance and applying a clustering algorithm such as K-means\cite{kanungo2002efficient}.

\subsubsection{Code Clone Detection}
There are supervised and unsupervised approaches to detect clones. While deep learning methods are applied to detect code clones, they require labelled data to train a supervised learning model~\cite{saini2018oreo, fang2020functional,zhang2019novel}. As such, one needs human annotators to mark pairs of snippets as clones, limiting the ability to detect clones by the amount of the data one can collect.

To alleviate the need of having labelled pairwise data to train supervised clone detector, 
we opt to use the unsupervised approach based on a similarity measurement:
For a pair of code snippets, we measure the similarity of the two vectors for the pair by using the cosine similarity;
when the cosine similarity between the vectors are higher than a certain threshold, we treat the pair as clones.
In this work, we choose 0.8 as the threshold.\lx{TODO: lack of justification; better to say it's chosen according to...some reason...cite some papers.}

\subsubsection{Cross Language Code-to-Code Search}
\textit{Code-to-code} search is useful for developers to find other code in a large code base that is similar to a given code query. For example, a developer working on a task to migrate a sorting algorithm implemented in Java to another language (e.g., C\#) might want to see if there exists an implementation of the same sorting algorithm in C\#, instead of rewriting the code in C\# from scratch.
Existing code-to-code search engine such as Krugle, Facoy~\cite{kim2018facoy}, Aroma~\cite{luan2019aroma}, only consider the searching problem within one programming language. Considering the more challenging use case that enables code-to-code search across multiple languages, our pre-trained InferCode model can be useful.
The backbone of InferCode is ASTs, and we used the ASTs from SrcML because it is a combined vocabulary for the AST node types in five main-stream languages (Java, C, C++, C\# and Objective C). Our pre-trained model can receive SrcML AST structure of any code snippets within these 5 languages. Given a code snippet in one language as a query, we aim to retrieve other code snippets that are functionally similar to the given code snippet in other programming languages. Since all code snippets can be represented in the form of vector representations, this problem can be formalized as the nearest-neighbor query in the vector space.

\subsection{Fine-Tuning for Supervised Learning Tasks}

A paradigm to make a good use of
large amount of unlabelled data is \textit{self-supervised pretraining followed by a supervised
fine-tuning}~\cite{hinton2006fast,chen2020simple}, which reuses parts (or all) of a trained neural network on a certain task and continue to train it or simply using the embedding output for other tasks. Such fine-tuning processes usually have the benefits of (1) speeding up the training as one does not need to train the model from randomly initialized weights and (2) improving the generalizability of the downstream model even when there are only small datasets with labels.
 
\begin{figure}[!t]
	\centering
	\includegraphics[scale=0.22]{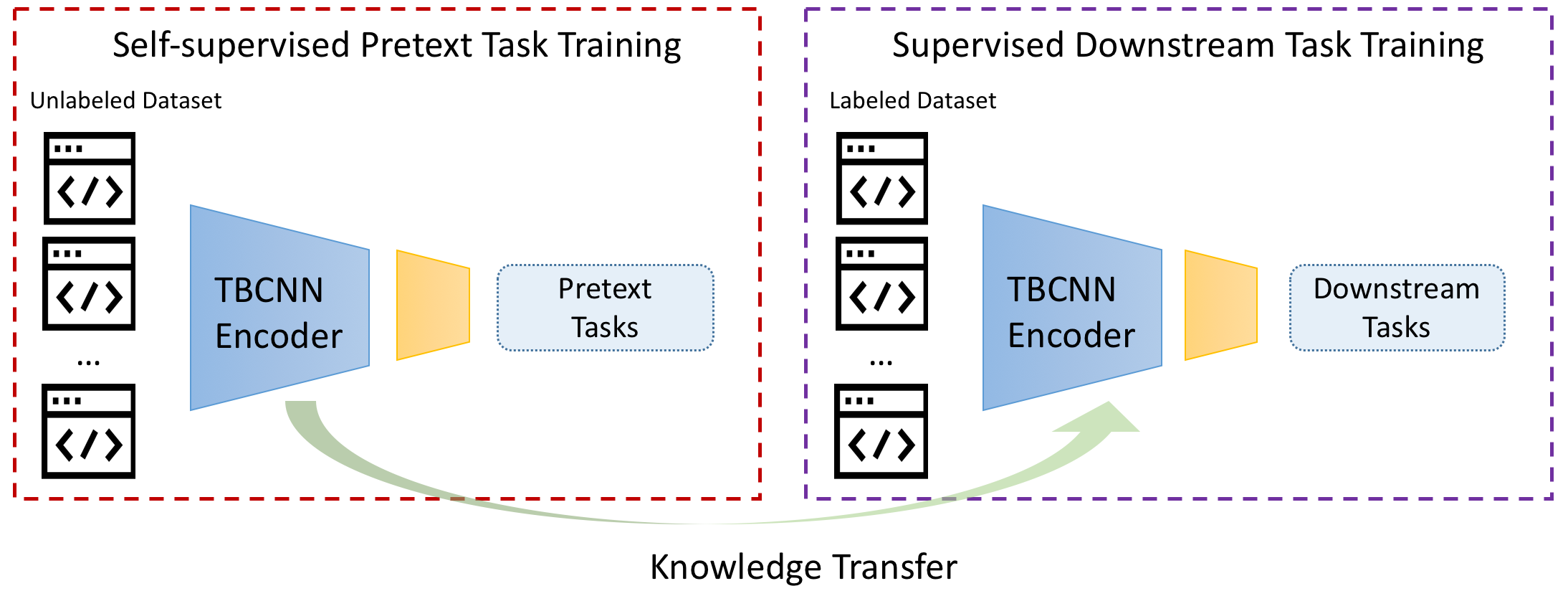}
	
	\caption{Code
		features are learned through the training process of TBCNN encoder to solve a predefined pretext task. After finishing the training, the learned parameters serve as a pre-trained model and can be transferred to other downstream tasks by fine-tuning. The performance on these downstream tasks is used to evaluate
		the quality of the learned features.}
	\label{fig:fine_tuning}
\end{figure}

As shown in Figure~\ref{fig:fine_tuning}, The TBCNN encoder of InferCode serves as a pretrained model, in which the weights resulted from the self-supervised learning are transferred to initialize the model of the downstream supervised learning task.

\subsubsection{Code classification}
We use \textit{code classification}~\cite{mou2016convolutional} as a downstream task to demonstrate the usefulness of the fine-tuning process.
This task is to, given a piece of code, classify the functionality class it belongs to.

\subsubsection{method name prediction}
We use \textit{Method name prediction}~\cite{Alon2019} as the second downstream task.
This task is to, given a piece of code (without its function header), predict a meaningful name that reflects the functionality of the code.
.

\section{Empirical Evaluation}\label{sec:evaluation}
In this section, we evaluate InferCode on the five use cases presented in Section~\ref{sec:use_case}.
We want to see to what degree the pre-trained model is applicable to different use cases even when the cases involve multiple programming languages.

To train our model, we reuse the Java-Large dataset that has been used in Code2vec \cite{Alon2019} and Code2seq \cite{alon2018code2seq}.
This dataset contains a large number of Java projects collected from Github (4 million files).
We parse all the files into ASTs using SrcML~\cite{collard2013srcml}. Then we identify all the subtrees to form a vocabulary of subtrees. Having the ASTs, and the subtrees as the pseudo labels, we train the InferCode model by using the softmax cross-entropy as the objective loss function and choose Adam \citep{kingma2014adam} as the optimizer with an initial learning rate of $0.001$  on an Nvidia Tesla P100 GPU. 

\subsection{Code Clustering}
\subsubsection{Datasets, Metrics, and Baselines}
We use two datasets for this task. The first is the OJ dataset that contains 52,000 C code snippets known to belong to 104 classes \cite{mou2016convolutional}. The second is the Sorting Algorithm (SA) dataset used in~\cite{Bui2019}, which consists of 10 classes of sorting algorithm written in Java, each algorithm has approximately 1000 code snippets.
Our clustering task here is to cluster all the code snippets (without class labels) according to the similarity among the code vectors: For the OJ dataset, 
we use K-means (K=104) to cluster the code into 104 clusters;
For the SA dataset, we use K-means (K=10) to cluster the code.
Then we use the class labels in the datasets to check if the clusters are formed appropriately.

We use the Adjusted Rand Index~\cite{santos2009use} as the metric to evaluate the clustering results.
Here we present the definition of Rand Index.
Let $C$ be the ground truth class assignment, and $K$ be the number of clusters assigned by a clustering algorithm.  Let $a$ be the number of pairs of elements that are in the same set in $C$ and the same set in $K$; and $b$ as the number of pairs of elements that are in different sets in $C$ and different sets in $K$. 
Rand Index for two datasets can be defined as:
$RI = \frac{a + b}{{n_{samples} \choose 2}}$, where the combinatorial number ${n_{samples} \choose 2}$ is the total number of possible pairs in the dataset (without ordering).
However, the $RI$ score does not guarantee that random label assignments will get a value close to zero (esp.~if the number of clusters is in the same order of magnitude as the number of samples). 
To counter this effect, {\em Adjusted Rand Index} is defined by discounting the expected $RI$ of random labelling as followed: 
$
ARI = \frac{RI - E[RI]}{max(RI) - E[RI]}
$.

For the baselines, if we treat source code as text, the self-supervised learning techniques in NLP can also be applied for code. As such, we include two well-known baselines from NLP, Word2vec~\cite{mikolov2013distributed}, and Doc2vec~\cite{le2014distributed}. We also include another baseline from~\cite{hill2016learning}, a state-of-the-art method to learn sentence representation. This method uses a Sequential Denoising Auto Encoder (SAE) method to encode the text into an embedding, and reconstruct the text from such embedding.
We also compare with two baselines for code modeling, Code2vec~\cite{Alon2019} and Code2seq~\cite{alon2018code2seq}. Code2vec works by training a path encoder on bag-of-paths extracted from the AST. The path encoder will encode the paths into an embedding $\vec{v}$, then use $\vec{v}$ to predict the method name. Code2seq shares a similar principle, but the $\vec{v}$ is used to generate text summary of code. In either case, we use the path encoders of Code2vec and Code2seq to produce the code vectors and also perform the same clustering process as InferCode.

\subsubsection{Results}
Table~\ref{tab:code_clustering} shows the results of code clustering using different models. InferCode performs the best for both datasets.
The NLP methods underperform other code learning methods.
This is reasonable because both Code2vec and Code2seq capture structural information from code, while NLP methods treat code as text sequences. We will provide a deeper analysis of the clusters by providing visualizations of the vectors produced by different methods (see Section \ref{sec:cluster_visualization}).
%
\begin{table}[!h]
	\centering
	\caption{Results of Code Clustering in Adjusted Rand Index (ARI)}
	\label{tab:code_clustering}
	\begin{tabular}{|c|c|c|}
		\hline
		\multirow{2}{*}{\textbf{Model}} & \multicolumn{2}{c|}{\textbf{Performance (ARI)}} \\ \cline{2-3} 
		& \textbf{OJ Dataset (C)} & \textbf{SA Dataset (Java)} \\ \hline \hline
		Word2vec                        & 0.28                & 0.24                \\ \hline
		Doc2vec                         & 0.42                & 0.29                \\ \hline
		SAE                             & 0.41                & 0.31                \\ \hline \hline \hline
		Code2vec                        & 0.58                & 0.51                \\ \hline
		Code2seq                        & 0.53                & 049                 \\ \hline \hline \hline
		InferCode                       & \textbf{0.70}                & \textbf{0.62}                \\ \hline
	\end{tabular}
\end{table}

\subsection{Code Clone Detection}
\subsubsection{Datasets, Metrics and Baselines}
We use two datasets in two languages. One is the OJ Dataset again that contains 52000 C programs. The other is the BigCloneBench, a Java dataset that has been widely used to benchmark code clone detection techniques, which consists of projects from 25,000 projects, cover 10 functionalities and including 6,000,000 true clone pairs and 260,000 false clone pairs.
For the OJ Dataset, we followed the process in~\cite{zhang2019novel} to construct a set of code pairs for clone detection based on pair-wise similarity measurement, so-called OJClone:
We choose 500 programs from each of the first 15 programming problems in OJ. It would produce a total of 1.8 million clone pairs and 26.2 million non-clone pairs, which are extremely time-consuming for comparison. So that we randomly select 50000 samples clone pairs and 50000 non-clone pairs for measuring the performance of various clone detectors.


We use the well-known Precision, Recall, and F1 scores.
Since the task is unsupervised, in this paper we compare InferCode only with unsupervised clone detectors that do not require labeled data (although the pretrained InferCode can also be applied to supervised clone detection).
The baselines include Deckard~\cite{jiang2007deckard}, SourcererCC~\cite{sajnani2016sourcerercc}, DLC~\cite{white2016deep}, and a detector using the code vectors extracted from Code2vec~\cite{Alon2019,kang2019assessing} and the same cosine similarity threshold used for InferCode.

\subsubsection{Results}
Table~\ref{tab:code_clone_detection} shows the overall precision, recall and
F1 for InferCode and other baselines. 
The detector based on InferCode has the highest recall (except for SourcererCC whose precision is relatively low).
Overall in terms of F1, it outperforms other
unsupervised clone detectors.

Note that we do not compare with techniques such as Oreo~\cite{saini2018oreo}, CCD~\cite{fang2020functional}, ASTNN~\cite{zhang2019novel} because they use supervised learning techniques to build clone {\em classifiers}.
We believe that the code embeddings or the weights from the pretrained InferCode can be used for training supervised clone classifiers too, and with further improvement on self-supervised learning techniques, such as improving the encoder, the auto-identified labels, and the loss function, the performance of unsupervised code clone detection may also get close to supervised ones. We leave these evaluations for future work.

\begin{table}[!h]
	\centering
	\caption{Results of Code Clone Detection in Precision, Recall and F1}
	\label{tab:code_clone_detection}
	\begin{tabular}{|c|c|c|c|c|c|c|}
		\hline
		\multirow{2}{*}{\textbf{Methods}} & \multicolumn{3}{c|}{\textbf{BigCloneBench (Java)}} & \multicolumn{3}{c|}{\textbf{OJClone (C)}} \\ \cline{2-7} 
		& \textbf{P}    & \textbf{R}   & \textbf{F1}   & \textbf{P} & \textbf{R} & \textbf{F1} \\ \hline \hline
		Deckard                           & 0.93          & 0.02         & 0.03         & 0.99       & 0.05       & 0.10        \\ \hline
		DLC                               & 0.95          & 0.01         & 0.01         & 0.71       & 0.00       & 0.00        \\ \hline
		SourcererCC                       & 0.88          & 0.02         & 0.03         & 0.07       & 0.74       & 0.14        \\ \hline
		Code2vec                          & 0.82          & 0.40         & 0.60         & 0.56       & 0.69       & 0.61        \\ \hline \hline
		InferCode                         & 0.90          & 0.56         & 0.75         & 0.61       & 0.70       & 0.64        \\ \hline
	\end{tabular}
\end{table}

\subsection{Cross Language Code-to-Code Search}
\subsubsection{Datasets, Metrics, and Baselines}
Given the implementation of an algorithm in one language, this task is to search for other implementations of the same algorithm written in other languages. So we need a dataset that contains multiple implementations of algorithms in different languages. We construct such a codebase for search from the Rosetta Code\footnote{\url{http://www.rosettacode.org}, \url{https://github.com/acmeism/RosettaCodeData}} and other code from GitHub:
We collect code in Java, C, C++, C\# from Rosetta Code which results in around 3000 samples;
then we collect 5000 random program files from Github for each of the languages and mix them with the samples.

For instance, for Java, we collect a large set of Java projects from Github that have at least 10 stars.
There is a possibility that the collected GitHub projects contain implementations of the algorithms in the Rosetta Code.
So we perform a simple text filtering to exclude all the files that contain a token of any of the algorithm name. Let us take 3 algorithms as examples (Bubble-sort, Singly-linked-list-Traversal, Yin-yang\footnote{These are taken from the names of the algorithms at \url{https://github.com/acmeism/RosettaCodeData/tree/master/Task}}): We exclude any file that contains any of these tokens: \textit{\{bubble, sort, singly, linked, list, traversal, yin, yang\}}. Then for the remaining Java files, we sample a subset of 5000 files and mix them with the Java implementations of the algorithms from the Rosetta dataset. We do the same for C\#, C++, C, so that we get in total about 23,000 files in our search codebase.

With the constructed code base, we perform the evaluation for cross-language search as follows:
For each of the 3000 code files from Rosetta Code, say a bubble sort implementation written in Java, we use it as the query to retrieve other files containing top-K similar code, we choose K = 10 in this evaluation. The ideal query results should only return a list of code snippets that are from Rosetta Code but implement the same bubble sort algorithm in C++, C\#, and C; other results would be considered as false positives. 
Since our assumption is that there is only one relevant result for the query, we use the well-known Mean Reciprocal Rank (MRR) as the metric to evaluate the actual query results.

Since this task can be formulated as the information retrieval (IR) problem and the neural IR techniques are widely applied recently for text data 
\cite{wang2011cascade,wan2016deep,vulic2015monolingual},
we include Word2vec, Doc2vec, CLIR~\cite{vulic2015monolingual}, a cross-lingual information retrieval system for text. We also follow ~\citet{Sachdev2018} to include ElasticSearch, a fuzzy text search baseline.
Although there are recent methods designed specifically for code-to-code search, such as Facoy \cite{kim2018facoy} and Aroma \cite{luan2019aroma}, they are designed only for monolingual code search, thus we do not compare with them directly. 

\subsubsection{Results}

Table~\ref{tab:code_search} shows the results for InferCode and other baselines. The performance of InferCode is the best among all the models. ElasticSearch, on the other hand, performs the worst; this is expected because ElasticSearch is a simple fuzz text search technique not designed to capture structural information of code. The performance of 
%
\begin{table}[!h]
	\centering
	\caption{Results of cross-language code-to-code search in Mean Reciprocal Rank (MRR)}
	\label{tab:code_search}
	\begin{tabular}{|c|c|c|c|c|}
		\hline
		\multirow{2}{*}{\textbf{Approach}} & \multicolumn{4}{c|}{\textbf{Performance (MRR)}}          \\ \cline{2-5} 
		& \textbf{Java} & \textbf{C\#} & \textbf{C++} & \textbf{C} \\ \hline
		ElasticSearch                      & 0.13          & 0.18         & 0.22         & 0.21       \\ \hline
		Word2vec                           & 0.33          & 0.36         & 0.30         & 0.32       \\ \hline
		Doc2vec                            & 0.32          & 0.34         & 0.38         & 0.30       \\ \hline
		CLIR                               & 0.29          & 0.32         & 0.34         & 0.39       \\ \hline \hline
		InferCode                          & \textbf{0.57}         & \textbf{0.45}         & \textbf{0.51}        & \textbf{0.54}       \\ \hline
	\end{tabular}
\end{table}

\subsection{Fine-Tuning for Supervised Learning Tasks}

\subsubsection{Datasets, Metrics, and Baselines}
\paragraph{Code Classification} We again use the OJ Dataset for this task. We split this dataset into three parts for training, testing, and validation by the ratio of 70:20:10.
Out of the training data, we feed X\% to the neural model, where X = 1, 10, 100. We then initialize the neural model either randomly or with the weights from the pre-trained InferCode. Therefore, we have four settings for training the supervised model for comparison: fine-tuning the TBCNN encoder with 1\%, 10\%, or 100\% of the labeled training data respectively, and the randomly initialized model.
Using only 1\% or 10\% is to demonstrate that given a pre-trained model, one only needs a small amount of labeled data to achieve reasonably good performance for the downstream task.

We use the accuracy metric widely used for classification tasks. As the baselines, we include the ASTNN \cite{zhang2019novel} trained from scratch, which is 
a state-of-the-art model for code classification on the OJ dataset, and
TextCNN~\cite{kim2014convolutional} and Bi-LSTM~\cite{schuster1997bidirectional} trained with 100\% of the training data, which are widely used for text classification.

\paragraph{Method Name Prediction} We use the Java-Small dataset widely used as a benchmark for method name prediction and has been used in Code2vec~\cite{Alon2019} and Code2seq~\cite{alon2018code2seq}. This dataset has already been split into three parts, namely training, testing, and validation.  We perform the same evaluation protocol as the code classification task by fine-tuning the model with 1\%, 10\%, and 100\% of the labeled training data, in contrast to random initialization of the model without fine-tuning.
To predict the method name, we follow Code2vec to use the code vector $\vec{v}$ to predict the embedding of a method name from a lookup table (see Section 4.2 in Code2vec~\cite{Alon2019}). We measure prediction performance using precision (P), recall (R), and F1 scores over the sub-words in generated names, following the metrics used by \citet{Alon2019}.
For example, a predicted name \texttt{result\_compute} is considered as an exact match of the ground-truth name \texttt{computeResult}; predicted \texttt{compute} has full precision but only 50\% recall; and predicted \texttt{compute\_model\_result} has full recall but only 67\% precision.

\subsubsection{Results} Table~\ref{tab:finetune_code_classification} shows the results for code classification.
Fine-tuning on 10\% of the training data gets comparable results with the NLP baselines. Fine-tuning on 100\% of the training data gets comparable with ASTNN, a state-of-the-art model for code classification on the OJ dataset.
\begin{table}[!t]
	\centering
	\caption{Results of Code Classification in Accuracy with Fine-Tuning (FT) on the OJ dataset}
	\label{tab:finetune_code_classification}
	\begin{tabular}{|p{1.5cm}|c|c|c|c|}
		\hline
		\textbf{Approach} & \textbf{FT (1\%)} & \textbf{FT (10\%)} & \textbf{FT (100\%)} & \textbf{Supervised} \\ \hline
		InferCode         & 70.4\%            & 87.6\%             & \textbf{98.0\%}              & 94\%                \\ \hline
		TextCNN             & -                 & -                  & -                   & 88.7\%              \\ \hline
		Bi-LSTM             & -                 & -                  & -                   & 88.0\%              \\ \hline
		ASTNN             & -                 & -                  & -                   & 97.8\%              \\ \hline
	\end{tabular}
\end{table}

Table~\ref{tab:finetune_method_name} shows the results for method name prediction. 
We get a comparable result with Code2seq when fine-tuning with 100\% labeled data.

\begin{table}[!t]
	\centering
	\caption{Result of Method Name Prediction in F1 with Fine-Tuning (FT) on the Java-Small Dataset}
	\label{tab:finetune_method_name}
	\begin{tabular}{|c|c|c|c|c|}
		\hline
		\textbf{Approach} & \textbf{FT (1\%)} & \textbf{FT (10\%)} & \textbf{FT (100\%)} & \textbf{Supervised} \\ \hline
		InferCode             & 20.31\%            & 30.54\%             & \textbf{43.33\%}              & 35.67\%                \\ \hline
		Code2vec             & -                 & -                  & -                   & 18.62\%              \\ \hline
		Code2seq             & -                 & -                  & -                   & 43.02\%              \\ \hline
	\end{tabular}
\end{table}


\subsection{Summary}

InferCode outperforms most of the baselines across five tasks, including three unsupervised ones (code clustering, code clone detection via similarity measurement), cross-language code-to-code search), and two supervised ones (code classification and method name prediction).

Note that this does not mean that the TBCNN encoder in InferCode is better than ASTNN, Code2vec, or Code2seq, as those neural models can be used as the encoder in InferCode too.
It only means that pre-training a model on large unlabeled data using self-supervised learning to predict subtrees can produce more transferable models while maintaining the performance of such models for various code learning tasks. 

The performance of the self-supervised learning models may be improved further with different encoders. We leave those explorations for future work.


\section{Analysis}
\label{sec:analysis}
This section analyses the effects of various parameters on the performance of different tasks.

\subsection{Cluster Visualization}
\label{sec:cluster_visualization}
\begin{figure*}[h]\centering
	
	\begin{tabular}{@{}cccc@{}}
		\includegraphics[width=0.66\columnwidth]{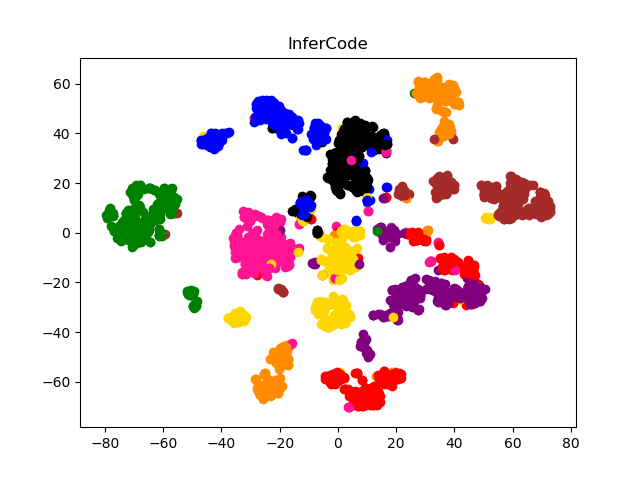} &
		\includegraphics[width=0.66\columnwidth]{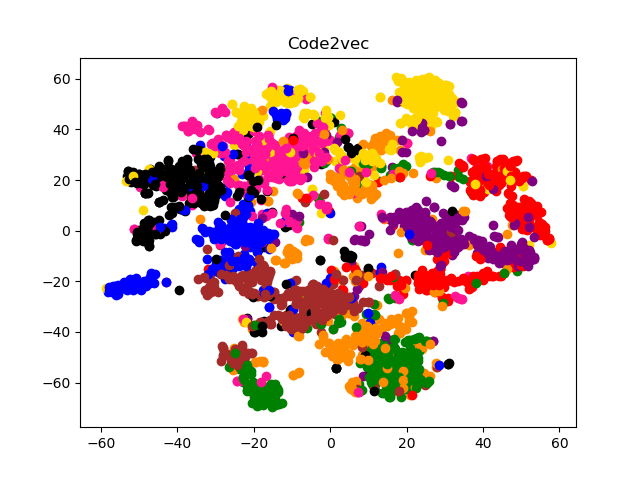} &
		\includegraphics[width=0.66\columnwidth]{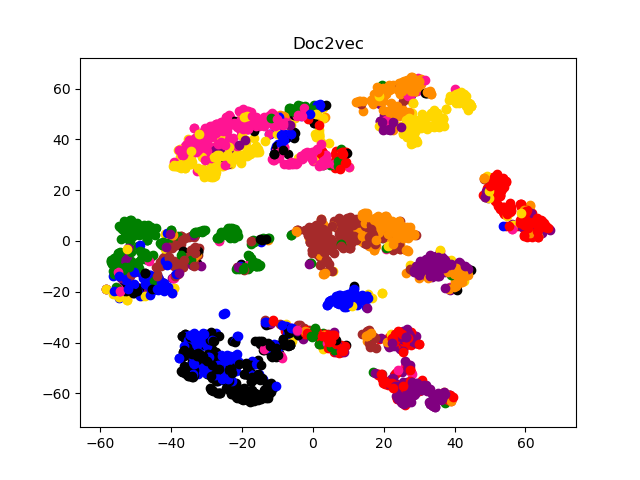}
	\end{tabular}
	\vspace*{-10pt}
	\caption{Visualization of the Code Vectors of the Programs from 9 classes in the OJ Dataset produced by InferCode, Code2vec and Doc2vec}
	\label{fig:visualization}
\end{figure*}

To help understand why the vectors produced by InferCode are better than the vectors produced by others, we visualize the vectors of the programs from the OJ dataset that have been used for the code clustering.
We choose the embeddings produced by Doc2vec, Code2vec, and InferCode for the first 9 classes of the OJ dataset, then we use T-SNE~\cite{maaten2008visualizing} to reduce the dimension of the vectors into two-dimensional space and visualize. As shown in Figure~\ref{fig:visualization}, (1) the vectors produced by InferCode group similar code snippets into the same cluster with clearer boundaries, and (2) The boundaries among clusters produced by Doc2vec and Code2vec are less clear, which makes it more difficult for the K-means algorithm to cluster the snippets correctly. This is aligned with the performance of the code clustering task (Table~\ref{tab:code_clustering}).
Also, we observe that some points marked in the same color (e.g., red) are somewhat far away from each other even in the vectors from InferCode, 
while they are supposed to be close according to the ground truth. This could indicate further improvements to Infercode can be made in future work.

\subsection{Effect of Textual Information in TBCNN}
\lx{this subsection is not very relevant for this paper; may remove to save space. Would be much better to perform an alation study on different encoders...}
The original TBCNN in \citet{mou2016convolutional} does not include textual information in AST nodes to initialize the node embedding. In our implementation, we include the textual information by fusing it with the node type information through a linear layer. To help understand the effect of such a fusion process, we perform an ablation study by training InferCode with different initialization information on the Java-Large dataset and perform the evaluations on the three unsupervised tasks: code clustering (CC), code clone detection (CCD), and cross-language code-to-code search (CLCS) with the same settings for each of the tasks in Section~\ref{sec:evaluation}.
Table~\ref{fig:effect_textual} shows the results of this study. Using only type or token information will result in worse performance for all three tasks.


\begin{table}[!h]
	\centering
	\caption{Effects of different initialization methods}
	\label{fig:effect_textual}
	\fontsize{7}{8}\selectfont 
	\begin{tabular}{|c|c|c|c|c|c|}
		\hline
		\multirow{2}{*}{\textbf{Task}} & \multirow{2}{*}{\textbf{Dataset}} & \multirow{2}{*}{\textbf{Metric}} & \multicolumn{3}{c|}{\textbf{Initial Information}}               \\ \cline{4-6} 
		&                                   &                                  & \textbf{Type} & \textbf{Token} & \textbf{Combine} \\ \hline
		CC                             & OJ                                & ARI                              & 0.57           & 0.28                 & \textbf{0.70}    \\ \hline
		CCD                            & BigCloneBench                     & P                               & 0.45           & 0.49                & \textbf{0.90}    \\ \hline
		CLCS                           & Rosetta Stone                     & MRR                              & 0.18           & 0.39                 & \textbf{0.57}    \\ \hline
	\end{tabular}
\end{table}

\subsection{Alternative Choices to the Pretext Task Labels}
There are a few alternatives when we use subtrees as the pseudo labels for the pretext task in InferCode.
One can easily replace the subtrees with tokens so that the code vector $\vec{v}$ can predict the tokens of the code snippets (similar to Doc2vec).
Or one can use all the method names as the pseudo labels and train the $\vec{v}$ to predict the names, similar to Code2vec~\cite{Alon2019}.
In this section, we perform an ablation study to measure how different types of labels can affect performance. 
As shown in Table \ref{fig:effect_prediction}, the performance using the subtrees as the labels is the best while using tokens as the labels result in the worst performance. Although using the method name can result in reasonable performance, it is still worse than using the subtrees. An explanation for this is that by predicting method names, the model is forced to learn some incorrect patterns due to similar names in the code base that actually refer to different code. For example, \citet{jiang2019machine} found that a large number code snippets contain similar method names but the actual implementations of the method bodies are different, but their code vectors would be forced to predict the similar method names, thus these vectors will be close in the vector space despite that they should not be. This is a potential reason to make the model trained by predicting method names a worse choice for pretext task than using subtrees.
\begin{table}[!h]
	\centering
	\caption{Effects of different ways to set up labels of the pretext task}
	\label{fig:effect_prediction}
	\fontsize{7}{8}\selectfont 
	\begin{tabular}{|c|c|c|c|c|c|}
		\hline
		\multirow{2}{*}{\textbf{Task}} & \multirow{2}{*}{\textbf{Dataset}} & \multirow{2}{*}{\textbf{Metric}} & \multicolumn{3}{c|}{\textbf{Label}}               \\ \cline{4-6} 
		&                                   &                                  & \textbf{Token} & \textbf{Method Name} & \textbf{Subtree} \\ \hline
		CC                             & OJ                                & ARI                              & 0.23           & 0.58                 & \textbf{0.70}    \\ \hline
		CCD                            & BigCloneBench                     & P                               & 0.45           & 0.81                 & \textbf{0.90}    \\ \hline
		CLCS                           & Rosetta Stone                     & MRR                              & 0.32           & 0.41                 & \textbf{0.57}    \\ \hline
	\end{tabular}
\end{table}

\section{Discussion}
\label{sec:disc}
\label{sec:discussion}
In this section, we want to discuss our choice on the decoder. We choose TBCNN because of its ability to capture structural features of code that lie in ASTs and the modification we made to TBCNN can also capture textual information into the model. There are many neural network designs that can be used as a replacement of the TBCNN encoder, such as ASTNN~\cite{zhang2019novel}, Code2vec~\cite{Alon2019} or GGNN~\cite{Allamanis2018}, however, most of them, especially the graph-based models, are unable to scale and generalize for different programming languages. For example, we can use the path encoder of Code2vec to encode the AST paths into the code vector $\vec{v}$ and infer the subtrees. GGNN is similar, one can pre-train the GGNN over a self-supervised learning task.
Although the graph representation proposed by~\citet{Narayanan2017,Allamanis2018} has been proved to work well on tasks, such as supervised clone detection, code summarization, variable name prediction, etc., choosing the suitable edges to be included in the graph representations for such tasks can be time-consuming and not generalizable.
LambdaNet~\cite{wei2020lambdanet} is another graph-based model that also contains semantic edges designed specifically for the type prediction task. As such, it is not straightforward to transfer a pre-trained graph learning model through different code learning tasks and it is not easy to scale the graph representation of code into multiple languages. Similar reasons can also be applied for path-based models, such as Code2vec and Code2seq, or execution trace-based models \cite{wang2019learning}.
On the other hand, TBCNN is designed to receive the AST directly with minimal engineering effort to process it.
AST is relatively easy to produce accurately for most programming languages given their grammars, thus building a tree-based learning model on top of ASTs implies that we can have a model that is easier to generalize across languages, which is the advantage to choose tree-based models over others.
Note that this is not to say that other models do not perform well on all the code learning tasks; they can still perform well when training data and time are specially utilized, and they may be used together with each other as the encoder in the self-supervised learning framework to improve the performance for various tasks further.
We leave all the exciting explorations for future work.


\section{Conclusions}
\label{sec:conclusion}
We have proposed InferCode, a self-supervised learning technique for source code learning on unlabeled data. The key intuition is that similar ASTs will have similar subtrees, which is aligned with the principle to learn document embeddings, where similar documents should contain similar words.  InferCode works by using a Tree-based CNN to encode the ASTs into a code vector and use it to predict the subtrees. We perform the training of InferCode on a large scale dataset. Then the encoder of InferCode, which is the Tree-based CNN can be reused as a pre-trained model. This pre-trained model is able to map the AST of any code snippet into an embedding and use it for other downstream tasks, such as code clustering, code clone detection, or code-to-code search. Our evaluation of these tasks show that the embeddings produce by InferCode are useful and outperform the other baselines with significant margins. Another use case of the pre-trained model is that its weights can be used under the notion of \textit{self-supervised pretraining followed by supervised fine-tuning}. We have shown that the fine-tuning process on a pre-trained model outperforms the supervised model trained from scratch. In the future, we will explore more on different choices of the encoder. We will also adapt InferCode into other tasks, such as bug localization, defect prediction, etc.

%

\balance
	\bibliographystyle{IEEEtranNurl}
	\bibliography{IEEEabrv,references}

\end{document}